\documentclass[12pt]{article}

\usepackage{graphicx}
\usepackage{mathptmx}      
\usepackage{epstopdf}
\usepackage{subfigure}
\usepackage{amsmath}

\newcommand{\mean}[2]{{\left< #1 \right>_{#2}}}
\newcommand{\uvec}[1]{\mathbf{\hat{#1}}}

\begin{document}

\title{Directionality of glioblastoma invasion in a 3d in vitro experiment.}
\author{A M Stein$^1$, D A Vader$^2$, T S Deisboeck$^3$, E A Chiocca$^4$, L M Sander$^5$ and D A Weitz$^2$ }

\maketitle

Corresponding author: \emph{amstein@umich.edu}

\begin{abstract}
Glioblastoma is the most malignant form of brain cancer.  It is extremely invasive; the mechanisms that govern invasion are not well understood.  To better understand the process of invasion, we conducted an \emph{in vitro} experiment in which a 3d tumour spheroid is implanted into a collagen gel.  The paths of individual invasive cells were tracked.  These cells were modeled as radially biased, persistent random walkers. The radial velocity bias was found to be 20 $\mu$m/hr on day one, but decayed significantly by day two.  The cause of this bias is thought to be due to chemotactic factors and contact guidance along collagen fibers.

\noindent{\it multicellular tumor spheroid, glioblastoma, invasion, Ornstein-Uhlenbeck, cell motility\/}
\end{abstract}

\section{Introduction}
The outcome for patients with highly malignant brain tumours is extremely poor.  Glioblastoma (GBM), the most malignant form of brain cancer, is responsible for 23\% of primary brain tumours and has a 5-year survival rate below 2.1\% \cite{CBTRUS03}.  One factor that makes GBM so difficult to treat is its high invasiveness \cite{Demuth04}.  It is known that the invasive cells are highly motile, but the mechanisms that govern their motility are not understood.  

In this paper, we present new results from experiments where fluorescently labeled tumour spheroids were grown in 3-dimensional collagen gels for two days.  These cells were modeled as radially biased random walkers using a 3-dimensional Ornstein-Uhlenbeck (OU) process.  The model fits the data and provides evidence for directed motility of the invasive cells away from the spheroid at a rate of 20 $\mu$m/hr on day one.  On day two, there is a dramatic decrease in the directional bias, which may be due to chemotactic factors or the way in which invasive cells reshape the collagen gel.

\section{Materials and methods}
\subsection{Cell culture}
U87MG glioblastoma cells with a modified EGF receptor \cite{Nishikawa94, Nagane96} (subsequently noted U87mEGFR) were cultured in Dulbecco's modified essential medium (DMEM) with 10\% fetal bovine serum (FBS), supplemented with 100 U/ml penicillin and 100 mg/ml streptomycin. The histone-GFP (H2B-GFP) fusion protein was stably expressed in U87mEGFR cells. The protein was transfected via the pBOS-H2BGFP vector, recently described by Kanda et al. \cite{Kanda98} and commercially available (BD Pharmingen, San Diego, CA). Blasticidin (VWR, Westchester, PA) at 5 $\mu$g/mL was added to the cell culture medium to select for cells expressing the fusion protein.

\subsection{Multicellular spheroids}
The cell suspension was subsequently diluted to a cell number density of $2.5\times10^4 $/mL and spheroids were produced using the hanging droplet method \cite{Kelm03}.  Briefly, 20 $\mu$L droplets of diluted cell suspension ($\sim$500 cells) were pipetted on the inside of the cover of a cell culture dish; the cover was then turned right side up and placed on top of the culture dish; the droplets were allowed to hang for 3 days until cells accumulated at the bottom of the drop through gravity and form a cell spheroid.

\subsection{Tumor model}
The extracellular matrix was modeled, \emph{in vitro}, by using type I
bovine collagen (Angiotech Biomaterials, Palo Alto, CA) at a final
concentration of 1.5 mg/mL. The collagen solution was prepared so as to
contain 10\% 10X DMEM, 10\% FBS, 1\% PS and 0.025 M Na$_2$HCO$_3$ (Invitrogen, Carlsbad, CA). A neutral pH was achieved by adding NaOH 1 M to the collagen solution.  The spheroid droplet was added after three days of hanging to 400 $\mu$L of collagen solution. The sample was then placed in an incubator (37$_\circ$C, 5\% CO$_2$, 100\% humidity) for 1 hour to allow collagen to polymerize. Finally, 100 $\mu$L of cell media was added on top of the polymerized sample to prevent drying.

\subsection{Image Acquisition}
\begin{figure}
		\centering
		\includegraphics[width=3in]{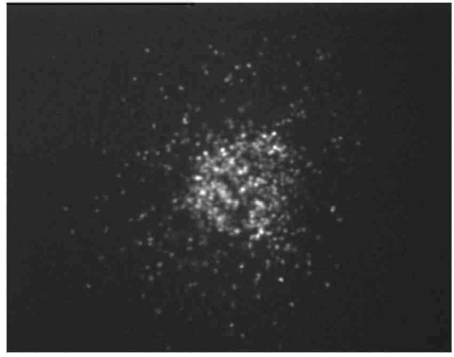}
		\caption{Wide field confocal microscopy image of the invasive tumour spheroid.  The image dimensions are 1700 x 1300 microns. \label{fig:expt_image}}
	\end{figure}

We used a Leica  DM IRB inverted microscope (Leica Microsystems, Wetzlar, Germany) with a 5X Leica objective to image our samples through a 640x480 Hamamatsu C7190 high sensitivity digital video camera (Hamamatsu Photonics, K. K., Hamamatsu City, Japan). The sample was placed on a heating stage (digital
tempcontrol 37-2, Leica) and covered with a CO$_2$ control chamber
(digital CTI controller 3700, Leica). We used a traditional FITC cube
with a 100W Mercury arc lamp (Ludl Electronic Products, Hawthorne, NY) to
excite the GFP-labeled nuclei and acquire fluorescent images of our sample.  A cell was visible if it was within approximately $25 \mu$m of the focal plane.  An example of an image taken from the setup is shown in figure~\ref{fig:expt_image}. 

After seeding the spheroid, we left the sample in the incubator for 6-8 hours, after which we imaged it every minute for 12 hours. The system was then placed in incubation again for 12 hours. Before imaging on the following day, cell media above the collagen gel was replaced by fresh cell media to ensure plentiful nutrient supply. Another sequence of images was then taken each minute for 15 hours.  From the images, the initial spheroid radius was determined to be 212 $\mu$m in radius. The coordinates for the center of the spheroid were obtained by eye.

\subsection{Cell tracking}
\begin{figure}
		\centering
		\subfigure[Cell paths from experiment: day 1]
			{\includegraphics[width=2.3in]{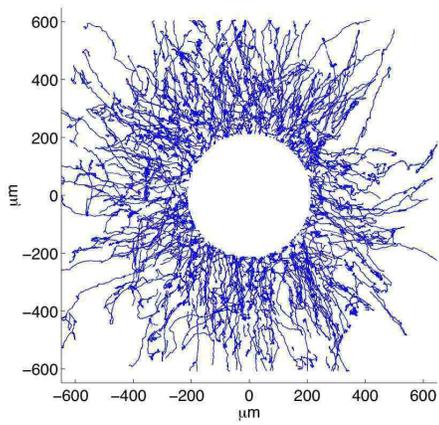}}
		\subfigure[Cell paths simulated by (\ref{eq:OU}): day 1]
			{\includegraphics[width=3in]{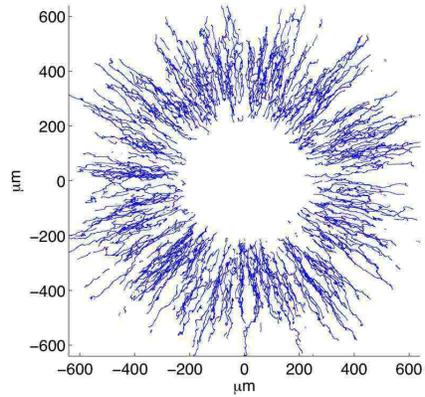}}
		\subfigure[Cell paths from experiment: day 1 (inner paths) and day 2 (outer paths)]
			{\includegraphics[width=3in]{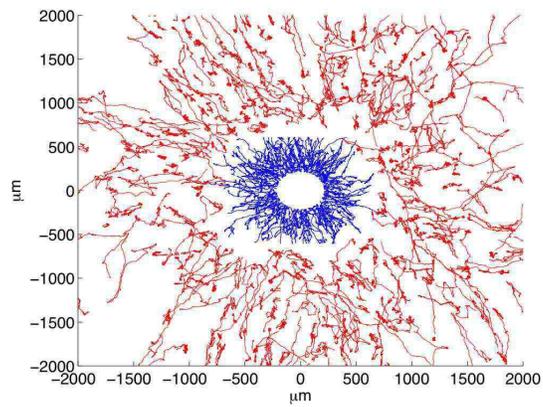}}
		\caption{Cell paths from experiment and from Ornstein-Uhlenbeck model. \label{fig:expt_paths}}
	\end{figure}

Identifying the nuclei on the images was achieved by a particle tracking method previously described by Crocker and Grier \cite{Crocker96} and implemented in the IDL (Research Systems, Inc, Boulder, CO) programming language. The output of the tracking routine is particle position, velocity, brightness and radius for each time frame, as well as a particle ID assigned automatically by the program. All subsequent analysis of cell trajectories was done using in-house written Matlab (Mathworks, Inc. Natick, MA) code.  Note that due to the high density of cells in the tumour core, the tracking algorithm produces errors there.  Thus all tracks located at the center of the image in the region estimated to be the core were ignored.  It should be noted that the core grew less than 60 microns over the course of two days, whereas cells invaded more than 2 mm; thus core growth was considered negligible. The paths are shown in figure~\ref{fig:expt_paths}; note that cells followed on day two are not necessarily those tracked on day one.  Because our analysis summarizes the behavior of the entire cell population, following different cells on day 2 should not effect the results.


\section{Mathematical Model}	
A model that is frequently used to describe cell motility is the continuous, persistent random walk, described by the Ornstein-Uhlenbeck equation \cite{Dunn87,Stokes91,Ionides04}, shown below in spherical coordinates $(r,\theta,\phi)$:
	\begin{align}
		\nonumber d V_r & =  -\beta_r(V_r - V_0 )dt & + \alpha_r d W \\
		d V_\theta & =  -\beta_\theta V_\theta dt & +  \alpha_\theta d W \label{eq:OU} \\
		\nonumber d V_\phi & =  -\beta_\phi V_\phi dt & + \alpha_\phi d W
	\end{align}
Here, $V$ is the velocity and $W$ is the Wiener process that represents the unpredictable aspects of cell motion.  We also introduce a velocity bias, $V_0$ in the radial direction away from the tumor spheroid.  As will be seen in section \ref{sec:Results}, this bias is necessary for describing the data.  One can extract physical meaning from $\alpha$ and $\beta$ by noting that the persistence time $P$ is equal to $1/\beta$, and the diffusion constant, $D$ is equal to $n\alpha/\beta^2$, where $n$ is the number of dimensions \cite{Stokes91}. 

The tumour spheroid was modeled as a 3d absorbing sphere 212 $\mu$m in radius to match the size of the spheroid in the experiment.  Each individual invasive cell was modeled as a 3d random walker that obeyed (\ref{eq:OU}).  At time 0, random walkers were uniformly distributed in a spherical shell with an inner radius of 212 $\mu$m and an outer radius of 412 $\mu$m. Their initial velocity was chosen to be radial and uniformly distributed in the interval $[0,2V_0]$.   To generate model paths to compare to experiment, we examined only the portion of each path that remained in a 50 $\mu$m planar slice that passed through the center of the spheroid.  This reflected the fact that cells could leave and enter the field of view of the microscope.

It should be noted that while cells were shed from a sphere that was over 400 microns in diameter, the slice in which the cells were observed was only 50 microns thick.  The 2d measurements thus allow us to observe $V_r$ and $V_\theta$, but not $V_\phi$.  It will be shown in the next section that on average, $V_\theta = 0$, and if we assume the system is spherically symmetric, so is $V_\phi$.  Thus we believe reducing our analysis to the 2D projection of the 3D system is reasonable.

\section{Results and discussion}
\label{sec:Results}
\begin{figure}
		\centering
		\includegraphics[width=4in]{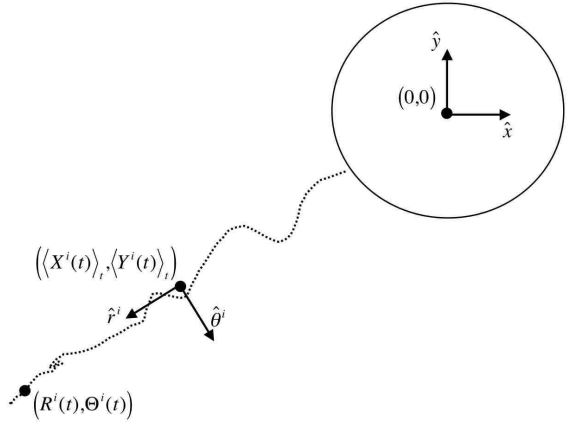}
		\caption{Global and local coordinate system for cell path $i$. \label{fig:coord}}
	\end{figure}

The paths from the experiment were analyzed so that the parameters for the OU model could be estimated. The horizontal and vertical position recorded for each cell, indexed by $i$, is $X^i(t)$ and $Y^i(t)$ respectively, where (0,0) is the center of the spheroid.  Note that although the cells moved in 3d, only the 2d projection of their position was recorded.  We transform these positions by a rotation into a local coordinate system for each cell, giving radial position $R^i(t)$ and angular position $\Theta^i(t)$.  Note that $\Theta^i(t)$ is not an angle, but rather the distance the cell has traveled perpendicular to the radial direction. The local coordinate system $(\uvec{r}^i,\uvec{\theta}^i)$ is shown in figure \ref{fig:coord} and given in the equations below, where $X^i(t)$ is written as $X^i$ for clarity. 
	\begin{align*}
		\uvec{r}^i &= \frac{\mean{X^i}{t}}{\mean{R^i}{t}}\uvec{x} + \frac{\mean{Y^i}{t}}{\mean{R^i}{t}}\uvec{y} \\
		\uvec{\theta}^i &= -\frac{\mean{Y^i}{t}}{\mean{R^i}{t}} \uvec{x} + \frac{\mean{X^i}{t}} {\mean{R^i}{t}} \uvec{y} 
	\end{align*}

\begin{figure}
		\centering
		\includegraphics[width=4in]{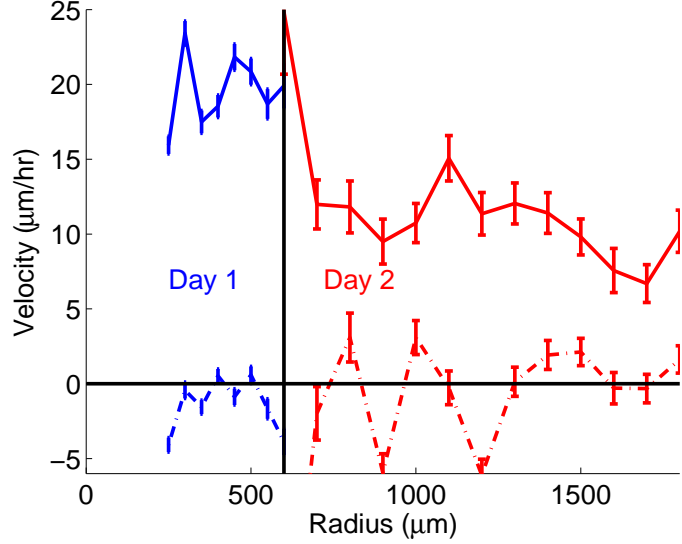}
		\caption{The radial velocity $V_r(a)$ (solid line) and angular velocity $V_\theta (a)$ (dashed line) at day one ($R \le 600 \mu m$) and day two ($R > 600 \mu m$) ensemble averaged over all cells and time as a function of distance from the center of the spheroid. \label{fig:vavg}}
	\end{figure}

Throughout this report, the notation $\mean{X}{t}$ indicates the variable $X$ averaged over $t$.  The velocity of a cell in the radial direction $V^i_r$, and angular direction $V^i_\theta$, is given by a forward difference, where $h$ is five minutes:  
	\begin{eqnarray*}
		V^i_r(t) &= \frac{R^i(t+h)-R^i(t)}{h} \\
	  	  V^i_\theta(t) &= \frac{\Theta^i(t+h)-\Theta^i(t)}{h}
		 \label{eq:vel}
	\end{eqnarray*}
To estimate the velocity bias $V_0$ from the experiment, we observe how the average radial velocity changes with the radial distance from the center of the spheroid.  This quantity, $\mean{V_r}{i,t}(a)$, also written $V_r(a)$, is estimated, as shown in (\ref{eq:vavg}), by averaging the velocities of all the cells that fall in a ring of thickness $\Delta a =$ 20 $\mu$m that is located $a$ distance away from the center of the spheroid.  
	\begin{equation}
	V_r(a) = \frac{1}{N(a)}\sum_{i,t} \big\{ V_r^i(t) | R^i(t) \in [a,a+\Delta a] \big\}
		\label{eq:vavg}
	\end{equation} 
Here, $N(a)$ is the number of velocity measurements made in the ring of radius $a$.  The average angular velocity, $V_\theta (a)$ is found in the same way.  The average radial and angular velocity are graphed in figure~\ref{fig:vavg}.  From the figure, we see that on day one, $V_r(a)$ and $V_\theta(a)$ are approximately constant, with $V_r(a) = 20$ $\mu$m/hr and $V_\theta(a) \approx 0$.   At later times, farther from the tumor spheroid, $V_r(a)$ decreases while $V_\theta (a)$ remains close to zero.  In comparison, Chicoine and Silbergeld \cite{Chicoine95} observed speeds of 12 $\mu$m/hr in glioblastoma cells on 2d substrates and Werbowetski et. al. observed speeds of 16 $\mu$m/hr for C6 astrocytoma spheroids \cite{Werbowetski04}. 

\begin{figure}
		\centering
		\includegraphics[width=6in]{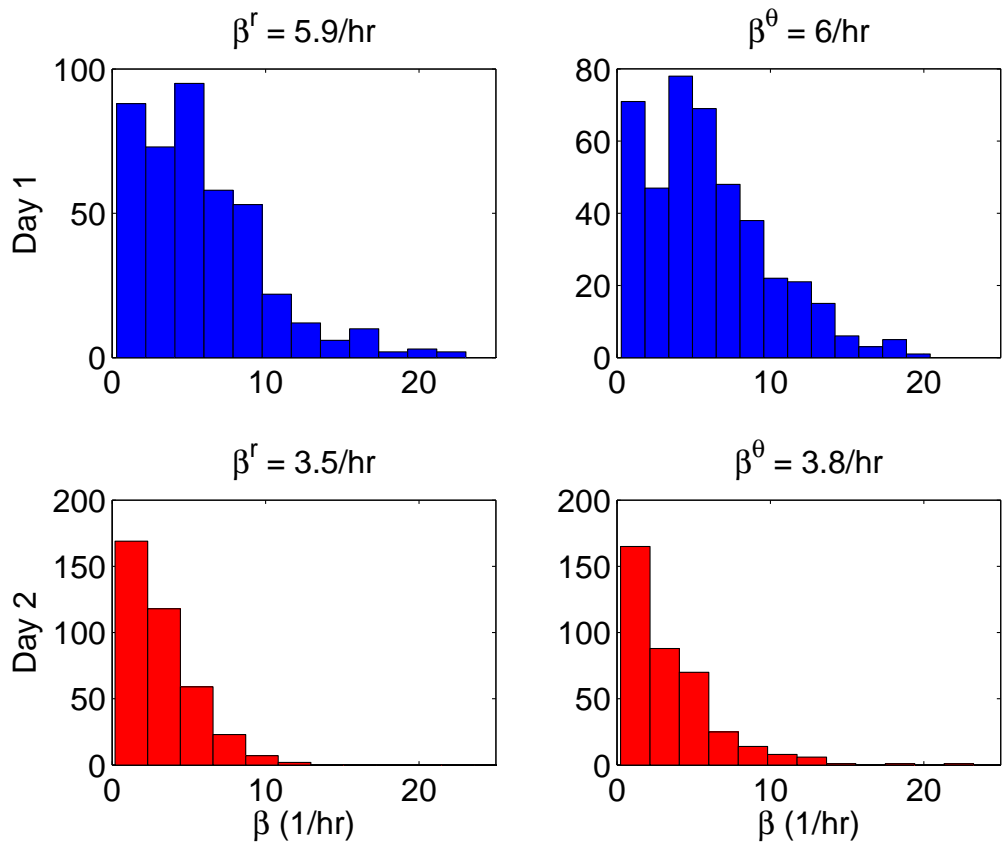}
		\caption{A histogram of $\beta_r$ and $\beta_\theta$ for all cells \label{fig:autocorr}}
	\end{figure}

For the OU process, the autocorrelation function of the velocity, $r(\tau)$, decays at a rate $e^{-\beta \tau}$.  Thus the autocorrelation function can be used to estimate $\beta_r$ and $\beta_\theta$.  We assume this system is spherically symmetric, and thus $\beta_\theta = \beta_\phi$.  We begin by finding the autocovariance function of an individual cell $c^i(\tau)$.  Note that the same equation applies for both $r$ and $\theta$ and we omit the subscript for clarity.
	\begin{equation}
		c^i(\tau) = \frac{1}{N^i(\tau)} \sum_t 
		(V^i(t)-\mean{V^i(t)}{t})(V^i(t-\tau)-\mean{V^i(t)}{t})	
		\label{eq:autocorr}
	\end{equation}
Here $N^i(\tau)$ is the number of pairs of points for a particular cell path that are spaced $\tau$ apart.  The autocorrelation for a particular cell is given by $r^i(\tau) = c^i(\tau)/c^i(0)$.  The distribution of $r^i(\tau)$ is shown in figure~\ref{fig:autocorr}, and we see that on day 1, $\beta_r = 5.9$/hr and $\beta_\theta = 6$/hr.

\begin{figure}
		\centering
		\includegraphics[width=5in]{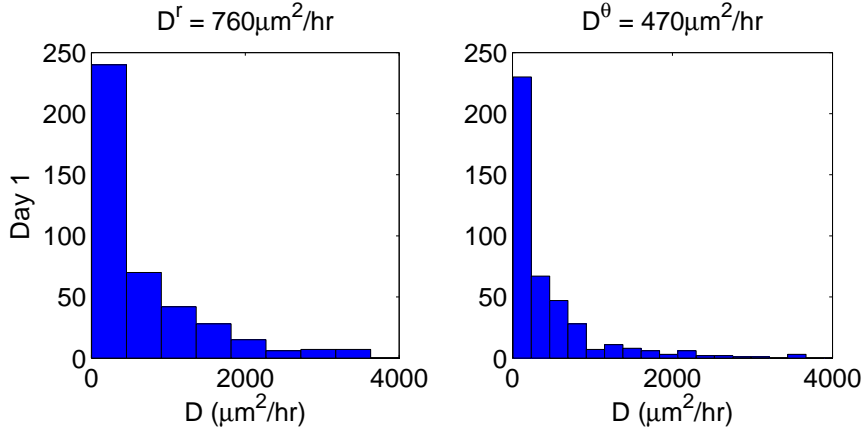}
		\caption{A histogram of the unbiased mean square displacement in the radial and angular dinections.  \label{fig:diff}}
	\end{figure}

To estimate the rate of diffusion of an unbiased random walker, one simply calculates the slope of the mean square displacement as a function of time.  In the case of a constant bias, as seen in the day 1 data, we estimate diffusion coefficients for a particular cell by first subtracting off the velocity bias, $\widetilde{R}^i(t) = R^i(t) - V_0^it$, where $V_0^i = \mean{V_r^i(t)}{t}$.  The unbiased mean squared displacement in the radial and angular directions is then estimated as follows:
	\begin{eqnarray*}
		\mean{(\widetilde{R}^i)^2}{t}(\tau) &= \frac{1}{N^i(\tau)} \sum_t
		\left( \widetilde{R}^i(t+\tau)-\widetilde{R}^i(t) \right)^2 \\
		\mean{(\Theta^i)^2}{t}(\tau) &= \frac{1}{N^i(\tau)} \sum_t
		\left( \Theta^i(t+\tau)-\Theta^i(t) \right)^2 \\
	\end{eqnarray*}
The diffusion coefficients, $D_r^i$ and $D_\theta^i$ are then estimated for each cell by the slope of the best fit least squares line through $\mean{(\widetilde{R}^i)^2}{t}(\tau)$ and $\mean{(\Theta^i)^2}{t}(\tau)$, and then averaged across all cells to get a population diffusion coefficient of $D_r = 760 \mu$m/hr and $D_\theta = 470 \mu$m/hr on day 1.  The results are shown in figure~\ref{fig:diff}.  We restricted the fits over the range $\tau \in [.3, 2]$ hours because at longer times intervals, there were fewer data points and plots of the mean square displacement at large $\tau$ often were not linear, as also seen in \cite{Stokes91}.  Using the formula $\alpha = 3 D \beta^2$ in three dimensions, we found $\alpha_r = 9120 \mu$m$^2$/hr$^3$ and $\alpha_\theta = 5640 \mu$m$^2$/hr$^3$.  On day 2, when there is no longer a constant radial velocity bias \emph{c. f.} figure~\ref{fig:vavg}, it was unclear how to estimate $D$.

\begin{figure}
		\centering
		\subfigure[Simulated cell paths]
			{\includegraphics[width=2in]{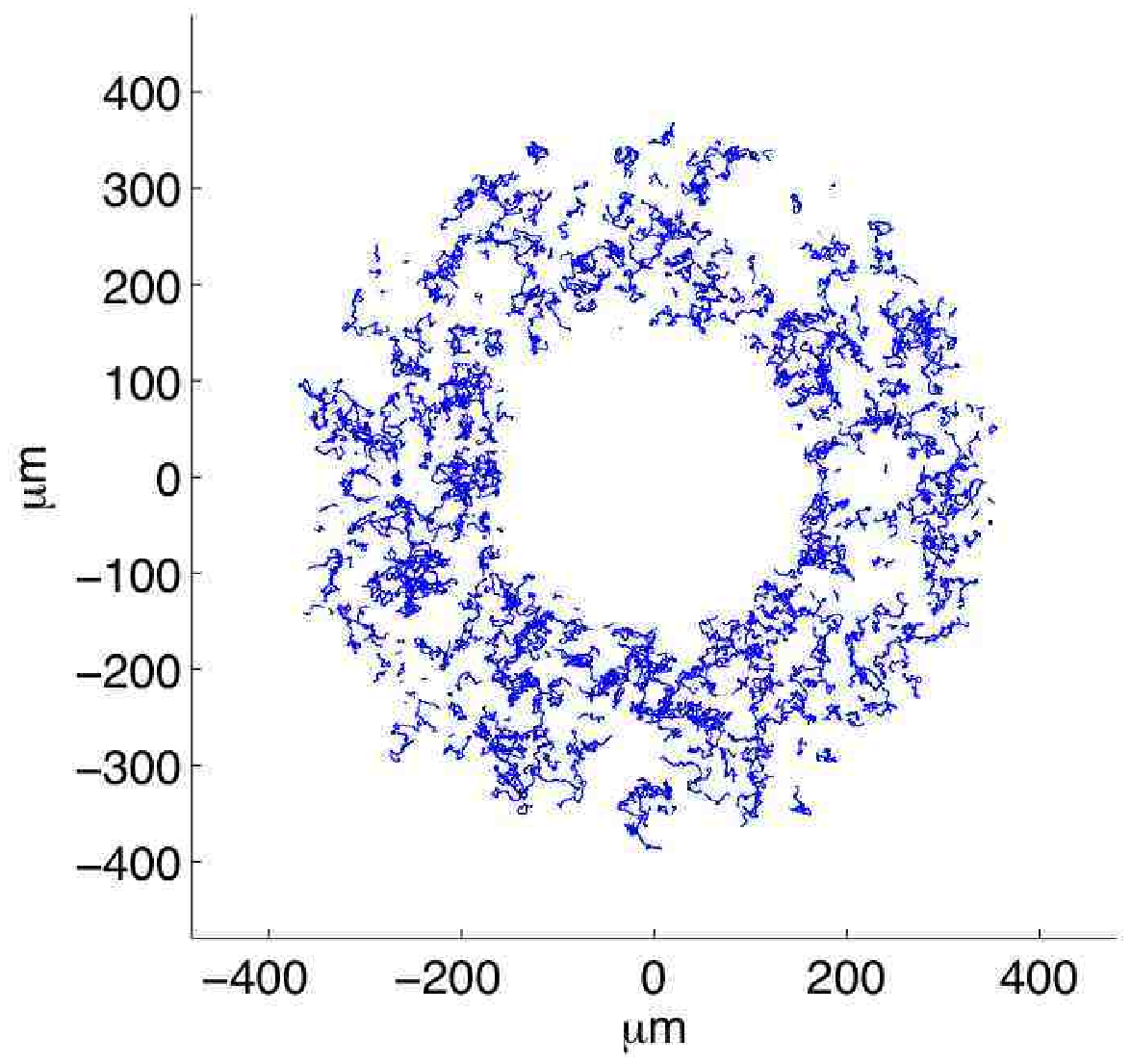}}
		\subfigure[Radial and Angular Velocity]
			{\includegraphics[width=2in]{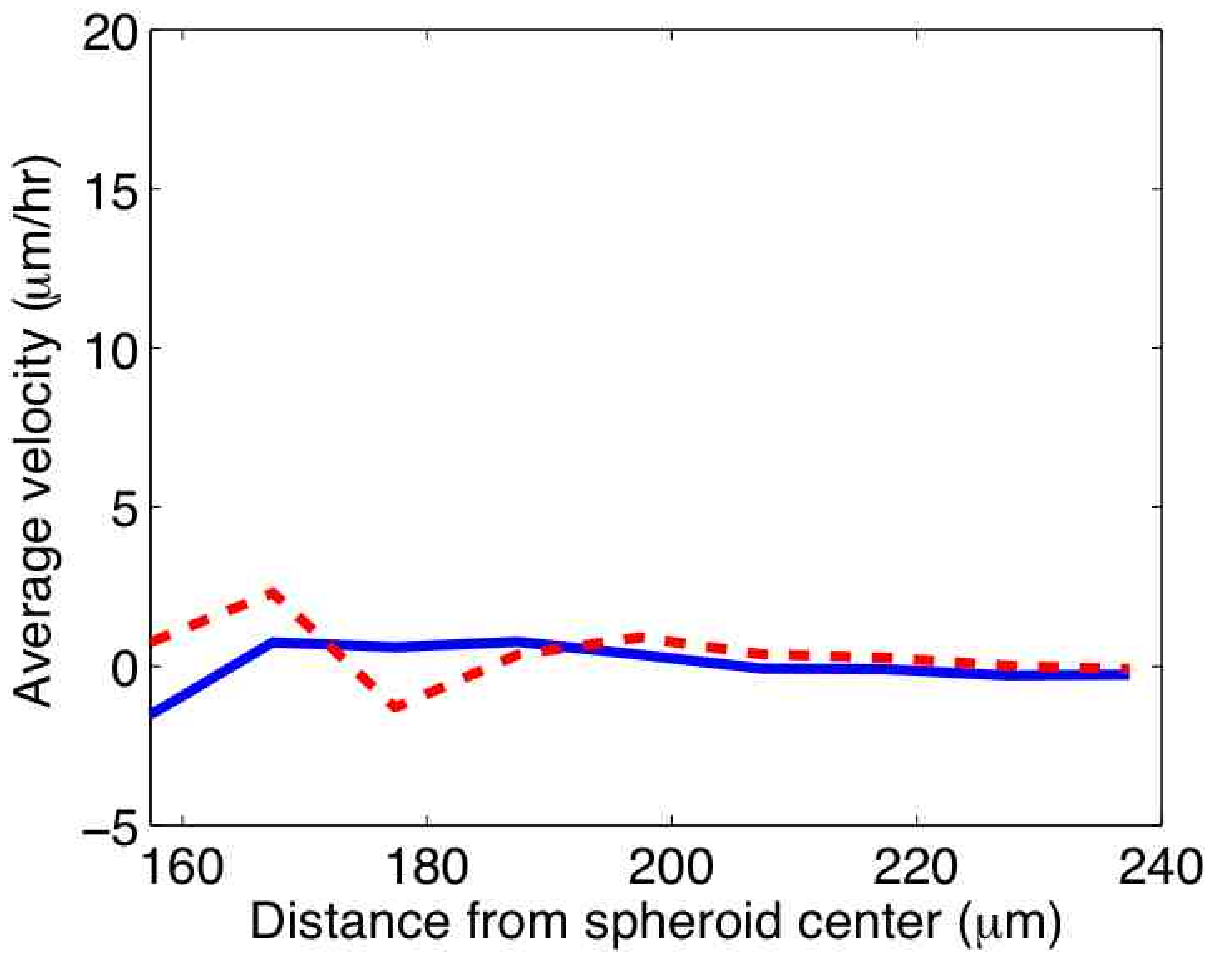}}
		\caption{Model simulation on day 1 with the $V_0 = 0$.  Shown here are the radial velocity (solid line) and angluar velocity (dashed line).  The radial bias that was present in the experiments is not seen here.  \label{fig:noV0}}
	\end{figure}

The model simulation with the parameters described above is shown in figure~\ref{fig:expt_paths} for day 1.  Of course, cells without a radial bias that start out concentrated in a region will also have a $V_r(a)$ that is greater than zero because cells will diffuse away from the spheroid with time.  To verify that this is not the cause of the bias, the model was run with the same parameter values for $\alpha$ and $\beta$ above, but with $V_0 = 0$.  The results are shown in figure~\ref{fig:noV0}.  
	
\section{Conclusions and outlook}
\begin{figure}
		\centering
		\includegraphics[width=4in]{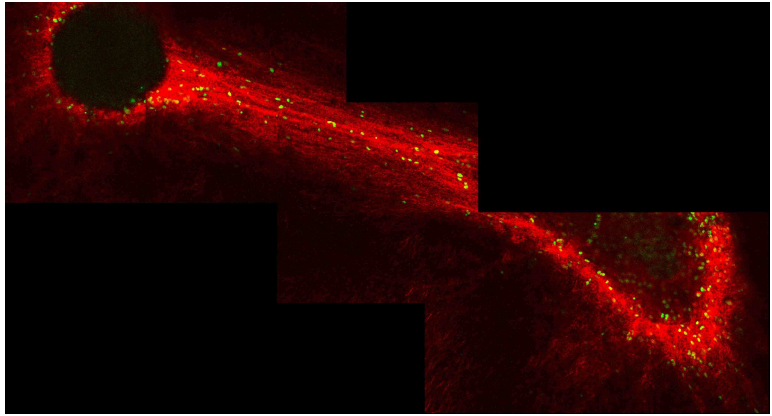}
		\caption{Two U87mEGFR spheroids were placed 2 mm from each other.  Two days later, restructuring of the collagen fibers is visualized (red) by confocal reflectance imaging.  U87mEGFR cells are transfected with a nuclear GFP so that the cell nuclei are labeled green.  Cells appear to migrate preferentially along the collagen fibrils between the two tumor spheroids. \label{fig:two_spheroids}}
	\end{figure}

The data provide clear evidence that when close to the spheroid, cells initially move away from it at a constant rate of 20 $\mu$m/hr and that as cells move away from the spheroid, the radial velocity bias decreases.  The cause of radial bias is unclear, though it is likely to be a combination of chemotaxis \cite{Werbowetski04, Eckerich05} and contact guidance on the reorganized collagen network \cite{Harris81, Kaufman05}.  The relative magnitude of these effects can be further investigated, for example by creating chemical gradients opposed to contact guidance and observing cell behavior. Additionally, restructuring of the collagen fibers can be more closely examined using confocal reflectance imaging, as shown in figure~\ref{fig:two_spheroids}.
It should be noted that there are differences between the simulated and real cell paths.  In the experiment, the path of an individual may have a short persistence time during one time interval and a long persistence time during a later time interval.  In our model, persistence time is constant.  It is not entirely clear what is causing the cell motility to change.  The periods of short persistence time may be due to cell-cell adhesion when cells are in close proximity or it may be due to cells staying in place during mitosis.  The cell motility is also likely influenced by the mesh size of the collagen fiber network.  It may also be that in isolation, a cell would naturally behave this way.  It is not yet clear if it is necessary to consider this phenomenon when describing invasion over longer time scales.

\section*{Acknowledgements}
We would like to thank T. Jackson, E. Ionides and E. Khain for helpful discussions.  Supported by NIH Bioengineering Research Partnership grant R01 CA085139-01A2.

\bibliographystyle{elsart-num}
\bibliography{BrainTumorLibrary}
\end{document}